\documentclass[pre]{revtex4}
\usepackage{amssymb}

\usepackage{eurosym}
\usepackage{amsmath}
\usepackage{graphicx}

\begin{document}

\title{Two- and one-dimensional gap solitons in spin-orbit-coupled systems
with Zeeman splitting}
\author{Hidetsugu Sakaguchi$^{1}$ and Boris A. Malomed$^{2,3}$}
\affiliation{$^{1}$Department of Applied Science for Electronics and Materials,
Interdisciplinary Graduate School of Engineering Sciences, Kyushu
University, Kasuga, Fukuoka 816-8580, Japan\\
$^{2}$Department of Physical Electronics, School of Electrical Engineering,
Faculty of Engineering, and Center for Light-Matter Interaction, Tel Aviv
University, Tel Aviv 69978, Israel\\
$^{3}$ITMO University, St. Petersburg 197101, Russia}

\begin{abstract}
We elaborate a mechanism for the formation of stable solitons of the
semi-vortex type (with vorticities $0$ and $1$ in their two components),
populating a finite bandgap in the spectrum of the spin-orbit-coupled binary
Bose-Einstein condensate with the Zeeman splitting, in the two-dimensional
free space, under conditions which make the kinetic-energy terms in the
respective coupled Gross-Pitaevskii equations negligible. Unlike a recent
work which used long-range dipole-dipole interactions to construct stable
gap solitons in a similar setting, we here demonstrate that stable solitons
are supported by generic local interactions of both attractive and repulsive
signs, provided that the relative strength of the cross/self interaction in
the two-component system does not exceed a critical value $\approx 0.77$. A
boundary between stable and unstable fundamental 2D gap solitons is
precisely predicted by the Vakhitov-Kolokolov criterion, while all excited
states of the 2D solitons, with vorticities $\left( m,1+m\right) $ in the
two components, $m=1,2,...$, are unstable. The analysis of the
one-dimensional (1D) reduction of the system produces an exact analytical
solution for the family of gap solitons which populate the entire bandgap,
the family being fully stable. Motion of the 1D solitons in the trapping
potential is considered too, showing that their effective mass is positive
or negative if the cubic nonlinearity is attractive or repulsive,
respectively.
\end{abstract}

\maketitle

\section{Introduction}

The realization of the settings which emulate the spin-orbit (SO)\ coupling
in ultracold atomic gases \cite{SOC1,SOC2,SOC-2D,SOC-review} has opened up a
vast research area in studies of matter waves. In particular, the interplay
of the linear SO coupling with intrinsic nonlinearity of Bose-Einstein
condensates (BECs) suggests many possibilities for the creation of vortices
\cite{SOC-vortices}, monopoles \cite{SOC-monopole}, skyrmions \cite%
{SOC-skyrmion}. Solitons have been predicted too, in both one-dimensional
\cite{soliton-1D} (including dark solitons in rinng-shaped condensates \cite%
{Brand}) and two-dimensional (2D) \cite{soliton-2D} geometries. A remarkable
fact is that the SO coupling makes it possible to stabilize 2D \cite{Ben
Li,Sherman} and three-dimensional (3D)\ \cite{Han Pu} solitons created by
the cubic attractive nonlinearity in the free space, while, in the usual 2D
and 3D models, solitons supported by cubic terms are always unstable because
of the presence of the critical (2D) and supercritical (3D)\ collapse \cite%
{old-review,Dumitru,recent-review}. Depending on the relative strength of
the self- and cross-attractive nonlinear interactions in the spinor
(two-component) SO-coupled BEC, the stable 2D and 3D solitons are \textit{%
semi-vortices}\ (alias \textit{half-vortices} \cite{Santos,Drummond-Pu}),
with one vortical and one zero-vorticity components, or \textit{mixed modes}%
, which combine vortical and zero-vorticity terms in both components \cite%
{Ben Li,Han Pu,Sherman}.

Recently, a new approach to the creation of stable 2D solitons in the
free-space BEC was proposed in Ref. \cite{rf3}. If the usual kinetic-energy
terms may be neglected in the respective system of Gross-Pitaevskii equation
(GPEs), the combination of the SO-coupling of the Rashba type \cite{Rashba}
and Zeeman splitting (which is a common ingredient of the SO-coupling
settings \cite{Zeeman}) replace the usual spectrum, determined by the atomic
mass, by one with a finite bandgap, which may be populated by families of
gap solitons, under the action of the attractive or repulsive nonlinearity.
The creation of gap solitons was first predicted \cite{Sterke} and
demonstrated experimentally \cite{Krug}, in an effectively 1D geometry, in
optical Bragg gratings, and later in exciton-polariton condensates loaded in
photonic lattices \cite{plasmon}. 1D gap solitons were also predicted \cite%
{BEC} and experimentally created \cite{Oberthaler} in the single-component
BEC\ trapped in an optical-lattice potential.

Getting back to the setting for the creation of 2D gap solitons proposed in
Ref. \cite{rf3}, it is relevant to note that, as demonstrated in that work,
if the model of the SO-coupled spinor BEC is effectively reduced from 3D to
2D by sufficiently tight trapping potential acting in the direction
perpendicular to the 2D plane, the kinetic-energy terms indeed turn out to
be negligible in comparison with the SO-coupling ones. Thus, the respective
2D\ system with the finite bandgap in the spectrum is quite a generic one.
It is relevant to mention that a possibility to neglect the kinetic-energy
terms in the SO-coupled BEC was demonstrated too in another setting, based
on a lattice potential which creates a \textit{flatband} in the respective
spectrum \cite{flatband}.

The two-component BECs which implement the SO coupling are usually realized
as mixtures of two different hyperfine atomic states. Accordingly, the
natural strengths of the self- and cross-interaction in this binary systems
may be close \cite{Ho}. Therefore, equal strengths of the contact self- and
cross-attraction were adopted in Ref. \cite{rf3}. The analysis has
demonstrated that such a model fails to create stable 2D gap solitons. For
this reason, long-range interactions were considered instead of the local
nonlinearity, assuming that the atoms in both states carry a permanent
magnetic dipole moment. This assumption is upheld by the prediction of the
realization of the SO coupling in a two-component dipolar BEC \cite{Cr}.
Furthermore, various effects which are possible due to the interplay of the
SO coupling and dipole-dipole interactions, were elaborated in a number of
works \cite{Xunda}. It has been demonstrated in Ref. \cite{rf3} that the
dipole-dipole interactions, both isotropic repulsive ones, in case the
dipole moments are polarized perpendicular to the system's plane, and
anisotropic interactions for the in-plane orientation of the moments \cite%
{Tikhonenkov}, readily create families of chiefly stable solitons populating
the entire system's bandgap. Moreover, the families extend as (partly
stable) \textit{embedded solitons} \cite{embedded} into a band across the
gap's top edge.

The possibility of the creation of \emph{stable} 2D gap solitons, and their
1D counterparts, making use of solely\emph{\ local} self- and
cross-interactions in the binary BEC is a challenging issue which is the
subject of the present work. An essential finding is that a subfamily of 2D
solitons of the semi-vortex type is stable, in the top part of the bandgap,
provided that the ratio of the strengths of the cross- and self-interaction
in the binary system takes values $\gamma <\gamma _{\max }\approx 0.77$, see
Eq. (\ref{max}) below. This finding explains the absence of stable 2D gap
solitons in Ref. \cite{rf3}, which only addressed the local nonlinearity
with $\gamma =1$. It is relevant to mention that $\gamma $ may be altered,
in broad limits, by means of the Feshbach-resonance technique, applying dc
magnetic field to the binary BEC \cite{Feshbach} (the long-range
dipole-dipole interactions can also be adjusted by means of experimentally
available methods \cite{Giovanazzi}). We find in this work that a boundary
between stable and unstable subfamilies of 2D semi-vortex solitons inside
the bandgap is precisely predicted by the well-known Vakhitov-Kolokolov
criterion \cite{VK,VK-review}. At the above-mentioned critical value of the
relative strength of the cross-interaction, $\gamma =\gamma _{\max }\approx
0.77$, the boundary merges into the top edge of the finite bandgap. It is
also relevant to stress that stable bright gap solitons can be supported
equally well by the contact interactions of either sign, attractive or
negative.

The rest of the paper is organized as follows. The model, based on coupled
2D GPEs and their 1D reduction, is formulated in Section II. The analysis
starts with the consideration of the reduced 1D system in Section III. A
full family of 1D gap solitons is found in an exact analytical form, and is
demonstrated to be completely stable. In addition to finding the solitons in
the free space, in the same section we consider motion of the 1D gap
solitons in an external trapping potential, concluding that the effective
soliton's mass is positive and negative if the contact nonlinearity is,
respectively, attractive or repulsive (in the latter case, the trapping
potential acts as an expulsive one). The above-mentioned findings for 2D gap
solitons of the semi-vortex type are summarized in Section IV. The paper is
concluded by Section V.

\section{The Gross-Pitaevskii equations}

The starting point is a scaled system of GPEs (with the atomic mass and $%
\hbar $ replaced by unity) for two components of the spinor wave function, $%
\Phi _{\pm }\left( X,Y,t\right) $ in the 2D space \cite{soliton-1D}-\cite%
{soliton-2D}:%
\begin{gather*}
i\frac{\partial \Phi _{+}}{\partial t}=-\frac{1}{2}\nabla ^{2}\Phi
_{+}+\lambda \left( \frac{\partial \Phi _{-}}{\partial X}-i\frac{\partial
\Phi _{-}}{\partial Y}\right) \Phi _{-}+\Omega \Phi _{+} \\
-\left( g|\Phi _{+}|^{2}+\gamma |\Phi _{-}|^{2}\right) \Phi _{+},
\end{gather*}%
\begin{gather}
i\frac{\partial \Phi _{-}}{\partial t}=-\frac{1}{2}\nabla ^{2}\Phi
_{-}-\lambda \left( \frac{\partial \Phi _{+}}{\partial X}+i\frac{\partial
\Phi _{+}}{\partial Y}\right) \Phi _{+}-\Omega \Phi _{-}  \notag \\
-\left( \gamma |\Phi _{+}|^{2}+g|\Phi _{-}|^{2}\right) \Phi _{-},
\label{Fulleq}
\end{gather}%
where {$g$ and }$\gamma $ are{\ coefficients of the self- and
cross-interactions of the components (their }positive and negative values
correspond, respectively, to attractive and repulsive interactions{),} while
$\lambda $ and $\Omega >0$ represent the strength of the SO coupling and
Zeeman splitting (the latter effect may be replaced by the Stark - Lo Surdo
splitting induced by dc electric field). These equations are derived from
the full 3D GPE system, assuming that the system is subject to tight
confinement in the transverse direction, with a trapping size $a_{\perp }$.
This derivation implies, as usual, that the 2D model is relevant for modes
(solitons) with lateral sizes $l\gg a_{\perp }$.

The SO-coupling effect in experimentally available settings is relevant if
it is strong enough, namely, $\lambda \gtrsim 1/a_{\perp }$, in the present
notation \cite{Drummond-Pu}. In the combination with condition $l\gg
a_{\perp }$, this inequality implies
\begin{equation}
l\lambda \gg 1.  \label{>>}
\end{equation}
Then, the ratio of the kinetic-energy and SO-coupling terms in Eq. (\ref%
{Fulleq}), obviously estimated as $\left( l\lambda \right) ^{-1}$, clearly
implies that the kinetic energy is negligible, which makes it possible to
reduce Eq. (\ref{Fulleq}) to the simplified form, without the second
derivatives and with $\left( X,Y\right) /\lambda \equiv $ $\left( x,y\right)
$, cf. Ref. \cite{rf3}:
\begin{eqnarray}
i\frac{\partial \Phi _{+}}{\partial t} &=&\left( \frac{\partial \Phi _{-}}{%
\partial x}-i\frac{\partial \Phi _{-}}{\partial y}\right) -(g|\Phi
_{+}|^{2}+\gamma |\Phi _{-}|^{2})\Phi _{+}+\Omega \Phi _{+},  \notag \\
i\frac{\partial \Phi _{-}}{\partial t} &=&-\left( \frac{\partial \Phi _{+}}{%
\partial x}+i\frac{\partial \Phi _{+}}{\partial y}\right) -(\gamma |\Phi
_{+}|^{2}+g|\Phi _{-}|^{2})\Phi _{-}-\Omega \Phi _{-}.  \label{2d}
\end{eqnarray}%
This is the basic model considered in the present work. If the nonlinear and
Zeeman terms are absent, Eq.~(\ref{2d}) is tantamount to the 2D equation for
Weyl fermions, which, in particular, finds its realization as a model of
graphene \cite{Weyl}. Equation (\ref{2d}) is also similar to 2D nonlinear
Dirac equations which were recently considered in different contexts \cite%
{Haddad1,Haddad2,Kevrekidis}, although in the latter works the cubic
nonlinearity is not sign-definite, unlike Eq. (\ref{2d}).

The 1D reduction of the system (when coordinate $y$ is absent) amounts to
the following equations:%
\begin{eqnarray}
i\frac{\partial \Phi _{+}}{\partial t} &=&\frac{\partial \Phi _{-}}{\partial
x}-(g|\Phi _{+}|^{2}+\gamma |\Phi _{-}|^{2})\Phi _{+}+\Omega \Phi _{+},
\notag \\
i\frac{\partial \Phi _{-}}{\partial t} &=&-\frac{\partial \Phi _{+}}{%
\partial x}-(g|\phi _{-}|^{2}+\gamma |\Phi _{+}|^{2})\Phi _{-}-\Omega \Phi
_{-}.  \label{1d}
\end{eqnarray}%
Strictly speaking, the reduction of the effective dimension from $2$ to $1$,
as well as from $3$ to $2$, or directly from $3$ to $1$, gives rise to
additional higher-order nonlinear terms; in particular, small effective
quintic terms are generated by the cubic ones in the underlying 3D GPE \cite%
{Luca,Maciek}, or in a system of coupled GPEs \cite{Luca2}. In the present
context, such additional terms are insignificant, as the simple cubic one
readily give rise to stable gap solitons, as shown below.

The linearization of Eqs. (\ref{2d}) and (\ref{1d}) for plane-wave
solutions, $\Phi _{\pm }\sim e^{ik_{x}x+ik_{y}y-i\mu t}$ gives rise to the
dispersion relation,
\begin{equation}
\mu =\pm \sqrt{\Omega ^{2}+k_{x}^{2}+k_{y}^{2}},  \label{gap}
\end{equation}%
and its 1D reduction (corresponding to $k_{y}=0$), which contains an obvious
gap, provided that $\Omega ^{2}$ is different from zero \cite{rf3}. As shown
below, the bandgap produced by this spectrum,%
\begin{equation}
\mu ^{2}\leq \mu _{\max }^{2}=\Omega ^{2},  \label{bandgap}
\end{equation}%
is completely populated by gap solitons, in 1D and 2D systems alike.

Before proceeding to the construction of the 1D and 2D gap solitons, it
makes sense to briefly discuss what may happen to them under the action of
small kinetic-energy terms dropped while deriving Eq. (\ref{2d}) from the
full system based on Eq. (\ref{Fulleq}). In terms of the notation which
keeps the SO-coupling strength, $\lambda $, as an explicit parameter, the
change of spectrum (\ref{gap}), caused by the presence of the second
derivatives in the full system, eventually closes the bandgap, at large
values of wavenumbers $k\sim \lambda $. In the course of extremely long
evolution, this will eventually lead to rearrangement of the gap solitons
into regular ones, through tunneling across the barrier separating the
(quasi-) gap and the range of $k\sim \lambda $. A straightforward estimate
demonstrates that, for generic solitons of width $l$, the tunneling time is
exponentially large with respect to parameter $l\lambda $, see Eq. (\ref{>>}%
). This estimate implies that the rearrangement time is many order of
magnitude larger than any realistic experimental time, hence the gap
solitons are completely robust objects, if Eq. (\ref{2d}) predict them as
stable solutions.

While the present work is chiefly focused on the consideration of quiescent
solitons, it makes sense to briefly consider a possibility to generate
moving ones too. This is a nontrivial issue, as Eqs. (\ref{2d}) and (\ref{1d}%
). although written in the free space, do not feature any apparent
invariance, such as Galilean or Lorentzian, which would allow one to
automatically generate moving solutions from quiescent ones. For this
purpose, it is appropriate to introduce moving coordinates, $\tilde{x}\equiv
x-c_{x}t,\tilde{y}\equiv y-c_{y}t$, where $\left( c_{x},c_{y}\right) $ is
the 2D velocity vector, and rewrite Eq. (\ref{2d}) in the moving reference
frame (cf. a similar approach developed in Ref. \cite{Ben Li}):%
\begin{gather}
i\frac{\partial \Phi _{+}}{\partial t}-i\left( c_{x}\frac{\partial \Phi _{+}%
}{\partial \tilde{x}}+c_{y}\frac{\partial \Phi _{+}}{\partial \tilde{y}}%
\right)   \notag \\
=\left( \frac{\partial \Phi _{-}}{\partial \tilde{x}}-i\frac{\partial \Phi
_{-}}{\partial \tilde{y}}\right) -(g|\Phi _{+}|^{2}+\gamma |\Phi
_{-}|^{2})\Phi _{+}+\Omega \Phi _{+},  \notag \\
i\frac{\partial \Phi _{-}}{\partial t}-i\left( c_{x}\frac{\partial \Phi _{-}%
}{\partial \tilde{x}}+c_{y}\frac{\partial \Phi _{-}}{\partial \tilde{y}}%
\right)   \notag \\
=-\left( \frac{\partial \Phi _{+}}{\partial \tilde{x}}+i\frac{\partial \Phi
_{+}}{\partial \tilde{y}}\right) -(\gamma |\Phi _{+}|^{2}+g|\Phi
_{-}|^{2})\Phi _{-}-\Omega \Phi _{-}.  \label{moving}
\end{gather}%
The linearization of Eq. (\ref{moving}) gives rise to the following
dispersion relation, cf. Eq. (\ref{gap}):%
\begin{equation}
\mu =\left( c_{x}k_{\tilde{x}}+c_{y}k_{\tilde{y}}\right) \pm \sqrt{\Omega
^{2}+k_{\tilde{x}}^{2}+k_{\tilde{y}}^{2}}.  \label{mu}
\end{equation}%
Simple analysis of Eq. (\ref{mu}) demonstrates that this spectrum contains a
reduced bandgap, provided that $c^{2}\equiv c_{x}^{2}+c_{y}^{2}<1$, with
width%
\begin{equation}
\mu ^{2}\leq \tilde{\mu}_{\max }^{2}=\sqrt{1-c^{2}}\Omega ^{2},
\label{reduced}
\end{equation}%
cf. Eq. (\ref{gap}). Edges of the reduced bandgap (\ref{reduced}) correspond
to wavenumbers%
\begin{equation}
\left( k_{\tilde{x}},k_{\tilde{y}}\right) =\mp \left( c_{x},c_{y}\right) /%
\sqrt{1-c^{2}}.
\end{equation}

\section{The one-dimensional system}

\subsection{Gap solitons in the free space}

We start the analysis with the 1D system, looking for stationary states with
chemical potential $\mu $ as $\Phi _{\pm }=e^{-i\mu t}\phi _{\pm }(x)$. The
substitution of this in Eq. (\ref{1d}) leads to equations for real wave
function $\phi _{\pm }(x)$:
\begin{eqnarray}
\mu \phi _{+}-\phi _{-}^{\prime }-\Omega \phi _{+}+(g\phi _{+}^{2}+\gamma
\phi _{-}^{2})\phi _{+} &=&0,  \notag \\
\mu \phi _{-}+\phi _{+}^{\prime }+\Omega \phi _{-}+(g\phi _{-}^{2}+\gamma
\phi _{+}^{2})\phi _{-} &=&0,  \label{U}
\end{eqnarray}%
with the prime standing for $d/dx$. Equations (\ref{U}) are similar to (but
different from) the system considered in Ref. \cite{we}, which was derived
as a model of a skewed dual-core optical waveguide.\ The exact solution for
1D gap solitons obtained in Ref. \cite{we} suggests a possibility to seek
for exact solutions to Eq. (\ref{U}). To this end, we note that evolution of
$\phi _{+}$ and $\phi _{-}$ along $x$ in the framework of Eq. (\ref{U})
conserves the corresponding Hamiltonian,
\begin{equation}
H=-\frac{\mu }{2}\left( \phi _{+}^{2}+\phi _{-}^{2}\right) +\frac{\Omega }{2}%
\left( \phi _{+}^{2}-\phi _{-}^{2}\right) -\frac{g}{4}\left( \phi
_{+}^{2}+\phi _{-}^{2}\right) ^{2}+\frac{g-\gamma }{2}\phi _{+}^{2}\phi
_{-}^{2}.  \label{h}
\end{equation}%
Then, looking for solutions in the polar form,
\begin{equation}
\phi _{+}(x)=A(x)\cos \left( \alpha (x)\right) ,~\phi _{-}(x)=A(x)\sin
\left( \alpha (x)\right) ,  \label{UUU}
\end{equation}%
and taking into account that solitons solutions, vanishing at $x\rightarrow
\pm \infty $, correspond to $H=0$, one can use Eq. (\ref{h}) to eliminate $%
A^{2}$ in favor of $\alpha $:
\begin{equation}
A^{2}=4\frac{-\mu +\Omega \cos (2\alpha )}{2g-(g-\gamma )\sin ^{2}(2\alpha )}%
.  \label{U^2}
\end{equation}%
Then, the substitution of expressions (\ref{UUU}) and (\ref{U^2}) in Eq. (%
\ref{U}) leads to an equation for $\alpha (x)$,
\begin{equation}
\alpha ^{\prime }=-\mu +\Omega \cos (2\alpha ),  \label{alpha'}
\end{equation}%
whose solution is easily found:
\begin{equation}
\alpha =\arctan \left[ \frac{\Omega -\mu }{\sqrt{\Omega ^{2}-\mu ^{2}}}\tanh
(\sqrt{\Omega ^{2}-\mu ^{2}}x)\right] .  \label{<0}
\end{equation}%
Note that, unlike the 2D system considered in the next section, which gives
rise to both fundamental gap solitons of the semi-vortex type and their
excited states (higher-order solitary vortices), the simple 1D equation (\ref%
{alpha'}) does not produce higher-order solitons.

The solution based on Eqs. (\ref{UUU})-(\ref{U^2}) and (\ref{<0}) yields the
family of 1D gap soliton with the chemical potential taking values in
bandgap (\ref{bandgap}). Here and in the next subsection, we consider the
gap solitons under the action of the dominant attractive interactions, which
means $g>0$ and $\gamma >-g$ (in particular, the case of $-g<\gamma <0$
corresponds to the competition of the self-attraction and weaker
cross-repulsion).

At the top edge of the bandgap, $\mu =+\Omega $, solution (\ref{<0})
degenerates into $\alpha \equiv 0$. Close to this edge, i.e., at $0<\Omega
^{2}-\mu ^{2}\ll \Omega ^{2}$, the solution takes the form of
\begin{gather}
\tan \alpha _{\mu \rightarrow \Omega }\approx \sqrt{\frac{\Omega -\mu }{%
2\Omega }}\mathrm{tanh}\left( \sqrt{2\Omega \left( \Omega -\mu \right) }%
x\right) ,  \notag \\
\left( \phi _{+}\right) _{\mu \rightarrow \Omega }\approx \frac{\sqrt{%
2\Omega \left( \Omega -\mu \right) }}{\cosh \left( \sqrt{2\Omega \left(
\Omega -\mu \right) }x\right) },  \notag \\
\left( \phi _{-}\right) _{\mu \rightarrow \Omega }\approx \left( \Omega -\mu
\right) \frac{\sinh \left( \sqrt{2\Omega \left( \Omega -\mu \right) }%
x\right) }{\cosh ^{2}\left( \sqrt{2\Omega \left( \Omega -\mu \right) }%
x\right) },  \label{vanishing}
\end{gather}%
which, in this approximation, does not depend on $\gamma $. The norm of the
limit-form gap soliton (\ref{vanishing}),%
\begin{equation}
N=\int_{-\infty }^{+\infty }\left[ \phi _{+}^{2}(x)+\phi _{-}^{2}(x)\right]
dx,  \label{N}
\end{equation}%
vanishes as $N_{\mu \rightarrow \Omega }\approx 2\sqrt{2\Omega \left( \Omega
-\mu \right) }$. At the bottom edge of the bandgap, $\mu =-\Omega $, the
solution remains nontrivial, being weakly localized, with $A^{2}(x)\sim
x^{-2}$ at $x\rightarrow \pm \infty $, rather than the exponentially
localized gap solitons found at $\mu ^{2}<\Omega ^{2}$:%
\begin{eqnarray}
\alpha _{\mu =-\Omega } &=&\arctan \left( 2\Omega x\right) ,
\label{edge-alpha} \\
A_{\mu =-\Omega }^{2}(x) &=&4\Omega \frac{1+4\Omega ^{2}x^{2}}{g+8\gamma
\Omega ^{2}x^{2}+16g\Omega ^{4}x^{4}}.  \label{edge-A}
\end{eqnarray}

While the entire family of the 1D gap solitons is found in the exact form,
as given by Eqs. (\ref{UUU})-(\ref{U^2}) and (\ref{<0})-(\ref{edge-A}),
their stability should be studied by means of numerical methods. We used
systematic simulations of perturbed evolution of the gap solitons to
identify their stability. Nevertheless, a necessary stability condition for
the gap-soliton family can be verified in a quasi-analytical form, if the
soliton's norm, defined as per Eq. (\ref{N}), is found as a function of $\mu
$ [in fact, we consider inverse dependences, $\mu (N)$]. Then, the necessary
stability condition for solitons supported by a dominant attractive
nonlinearity is provided by the well-known Vakhitov-Kolokolov (VK)
criterion, $d\mu /dN<0$ \cite{VK,VK-review}. It guarantees the absence of
imaginary eigenfrequencies in the spectrum of small perturbations around the
stationary soliton, which would give rise to an exponentially growing
instability; however, the VK criterion cannot detect complex
eigenfrequencies, that may generate an oscillatory instability.

While an analytical expression for the norm of the generic gap-soliton
solution, which corresponds to Eqs. (\ref{UUU})-(\ref{U^2}) and (\ref{<0}),
is too cumbersome to display it, the norm of the bandgap-edge soliton, given
by Eqs. (\ref{edge-alpha}) and (\ref{edge-A}), can be presented in an
explicit form:%
\begin{gather}
N\left( \mu =-1\right) =\pi \left[ \left( 1-\sqrt{\frac{\gamma -1}{\gamma +1}%
}\right) \sqrt{\gamma +\sqrt{\gamma ^{2}-1}}\right.  \notag \\
\left. +\left( 1+\sqrt{\frac{\gamma -1}{\gamma +1}}\right) \sqrt{\gamma -%
\sqrt{\gamma ^{2}-1}}\right] .  \label{mu=-1}
\end{gather}%
Here, to make the expression more compact, it is implied that $g=\Omega =1$
is fixed by means of scaling, and the expression is written for $\gamma \geq
1$. It is valid too for $\gamma <1$, as the analytical continuation, which
remains real and positive. Note that Eq. (\ref{mu=-1}) amounts to $N\approx
2\pi \sqrt{2/\gamma }$ at $\gamma \rightarrow \infty $.

\begin{figure}[h]
\begin{center}
\includegraphics[height=4.cm]{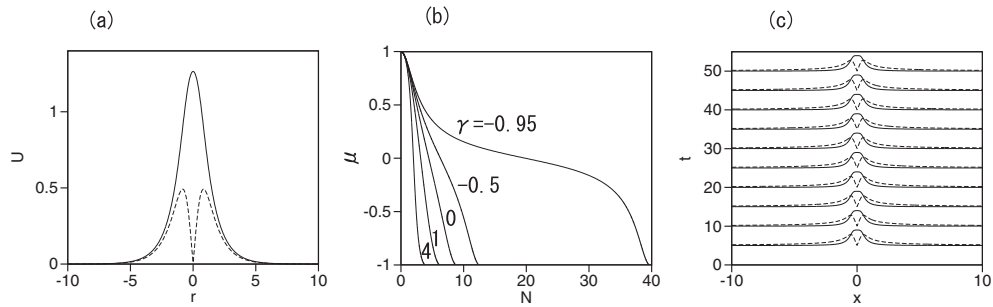}
\end{center}
\caption{(a) An exact 1D gap-soliton solution, given by Eqs. (\protect\ref%
{UUU})-(\protect\ref{U^2}) and (\protect\ref{<0}), for $\Omega =g=1,$ $%
\protect\gamma =0$, and $\protect\mu =0.2$. (b) Dependences $\protect\mu (N)$
(the total norm vs. the chemical potential) for the exact gap-soliton
families, for $\Omega =g=1$ and $\protect\gamma =-0.95$, $-0.5$, $0$, $1$,
and $4$. (c) The stable evolution of components $|\Phi _{+}|$ and $|\Phi
_{-}|$ of the bandgap-edge soliton, given by Eqs. (\protect\ref{edge-alpha})
and (\protect\ref{edge-A}), for $\protect\mu =-1$ at $\Omega =1$, $g=1$, and
$\protect\gamma =0$. Plot (c) is produced by simulations of Eq. (\protect\ref%
{1d}). }
\label{fig1}
\end{figure}

A typical example of a stable gap soliton is shown in Fig. \ref{fig1}(a) for
$\Omega =1$, $\mu =0.2$, $g=1$ and $\gamma =0$. Families of the gap
solitons, which entirely fill the bandgap, are represented by the
above-mentioned dependences, $\mu (N)$, in Fig. \ref{fig1}(b) for the same
values $\Omega =1$, $g=1$, and a set of different values of the
cross-interaction coefficient, $\gamma =-0.95,-0.5,0,1$, and $4$. As
mentioned above, $\gamma <0$ implies that the cross-interaction in Eq. (\ref%
{1d}) is repulsive, while fixed $g=1$ accounts for the self-attraction of
each component of the spinor wave function. It is worthy to note that the VK
criterion holds for all families of the 1D gap solitons at all values of $%
\mu $. In agreement with this fact, systematic simulations demonstrate that
the 1D gap solitons are completely stable. In particular, Fig. \ref{fig1}(c)
illustrates the stable evolution of a weakly localized bandgap-edge soliton,
whose norm is given by Eq. (\ref{mu=-1}).

Lastly, beneath the bandgap, i.e., at $\mu <-\Omega $, exact solutions to
Eq. (\ref{alpha'}) can be found too, in a spatially periodic form:%
\begin{equation}
\alpha =-\arctan \left( \sqrt{\frac{\mu -\Omega }{\mu +\Omega }}\tan \left(
\sqrt{\mu ^{2}-\Omega ^{2}}x\right) \right)  \label{periodic}
\end{equation}%
[at $\mu >\Omega $, i.e., above the bandgap, solution (\ref{periodic}) is
irrelevant, as its substitution in Eq. (\ref{U^2}) yields $A^{2}<0$].
However, unlike the solitons, these periodic patterns are unstable in direct
simulations (not shown here in detail).

\subsection{Dynamics of 1D gap solitons in the trapping potential}

Real experiments with BEC are always performed in the presence of trapping,
which is usually represented by the harmonic-oscillator potential, $%
U(x)=(1/2)kx^{2}$ with $k>0$. The accordingly modified 1D system of GPEs (%
\ref{1d}) is

\begin{eqnarray}
i\frac{\partial \Phi _{+}}{\partial t} &=&\frac{\partial \Phi _{-}}{\partial
x}-(g|\Phi _{+}|^{2}+\gamma |\Phi _{-}|^{2})\Phi _{+}+\frac{1}{2}kx^{2}\Phi
_{+}+\Omega \Phi _{+},  \notag \\
i\frac{\partial \Phi _{-}}{\partial t} &=&-\frac{\partial \Phi _{+}}{%
\partial x}-(g|\Phi _{-}|^{2}+\gamma |\Phi _{+}|^{2})\Phi _{-}+\frac{1}{2}%
kx^{2}\Phi _{-}-\Omega \Phi _{-}.  \label{1d2}
\end{eqnarray}

\begin{figure}[h]
\begin{center}
\includegraphics[height=4.cm]{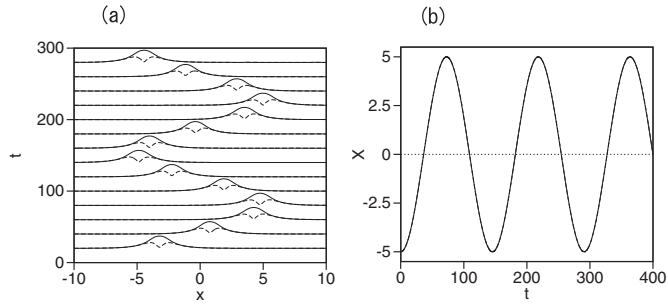}
\end{center}
\caption{(a) Continuous and dashed lines depict the evolution of $|\Phi _{+}|
$ and $|\Phi _{-}|$ for a 1D gap soliton which performs harmonic
oscillations in the trapping potential with $k=0.0025$, $\Omega =1$, $g=1$,
and $\protect\gamma =0$ in Eq.~(\protect\ref{1d2}). (b) The respective
evolution of the soliton's central coordinate, $X(t)$.}
\label{fig2}
\end{figure}

It is natural to expect that gap solitons may perform stable shuttle motion
in the trapping potential \cite{we-old} (then, the soliton's zero mode,
corresponding to the translational invariance in the free space, turns into
a Goldstone mode of small oscillations at the bottom of the potential well,
see, e.g., Ref. \cite{Goldstone}). This expectation is confirmed by
simulations of Eq. (\ref{1d2}). In particular, Fig. \ref{fig2}(a) displays
the time evolution of $|\Phi _{+}|$ and $|\Phi _{-}|$ of a gap soliton which
periodically moves in the trap with strength $k=0.0025$, and Fig. \ref{fig2}%
(b) shows the evolution of its center-of-mass coordinate,
\begin{equation}
X(t)=\frac{\int_{-\infty }^{+\infty }(|\Phi _{+}|^{2}+|\Phi _{-}|^{2})xdx}{%
\int_{-\infty }^{+\infty }(|\Phi _{+}|^{2}+|\Phi _{-}|^{2})dx},  \label{X}
\end{equation}%
initiated by initially placing the gap soliton at $X(t=0)=-5$. Accordingly,
the central coordinate exhibits a sinusoidal motion, $X=-5\cos (\omega t)$
with frequency $\omega =0.0432$. This result makes it possible to calculate
an effective mass of the gap soliton according to the elementary law of
harmonic oscillations,
\begin{equation}
m_{\mathrm{eff}}=k/\omega ^{2}\approx 1.34.  \label{meff}
\end{equation}%

\subsection{Gap solitons under the action of the repulsive interactions}

Repulsive nonlinearity can support gap solitons in the free space as well as
it is done by the self-attraction \cite{repulsion,Oberthaler,we-old}.
Indeed, starting from Eqs. (\ref{2d}) or (\ref{1d}) with $g<0$, one can
invert the sign of the nonlinearity coefficients, $g$ and $\gamma $, by the
substitution of $\Phi _{+}\rightarrow \Phi _{-}^{\ast }$, $\Phi
_{-}\rightarrow \Phi _{+}^{\ast }$, where the asterisk stands for the
complex conjugate. On the other hand, the self-repulsion makes the effective
mass of the gap soliton negative, hence one may expect that the trapping
potential will become expulsive for it \cite{we-old,negative-mass}. Here, we
address this point by simulating the 1D system in the form of
\begin{eqnarray}
i\frac{\partial \Phi _{+}}{\partial t} &=&\frac{\partial \Phi _{-}}{\partial
x}+(\tilde{g}|\Phi _{+}|^{2}+\tilde{\gamma}|\Phi _{-}|^{2})\Phi _{+}+\frac{1%
}{2}kx^{2}\Phi _{+}+\Omega \Phi _{+},  \notag \\
i\frac{\partial \Phi _{-}}{\partial t} &=&-\frac{\partial \Phi _{+}}{%
\partial x}+(\tilde{g}|\Phi _{-}|^{2}+\tilde{\gamma}|\Phi _{+}|^{2})\Phi
_{-}+\frac{1}{2}kx^{2}\Phi _{-}-\Omega \Phi _{-},  \label{1d4}
\end{eqnarray}%
with $\tilde{g}=1$, $\tilde{\gamma}\geq 0$ and $k>0$, which implies the
interplay of the repulsive nonlinearity and normal trapping potential.

\begin{figure}[h]
\begin{center}
\includegraphics[height=4.cm]{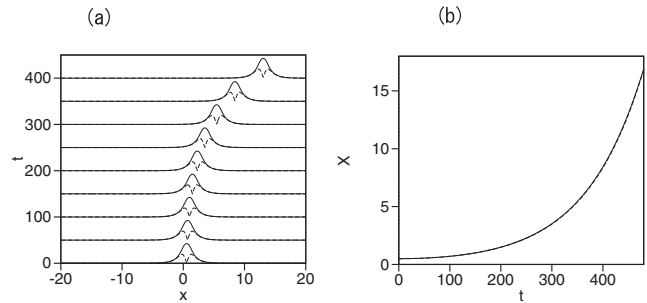}
\end{center}
\caption{The same as in Fig. (\protect\ref{fig2}), but in the case of the
interplay of the self-repulsive nonlinearity and trapping potential, as per
Eq. (\protect\ref{1d4}) with $k=0.0001$, $\Omega =-1,\tilde{g}=1$, and $%
\tilde{\protect\gamma}=0$.}
\label{fig3}
\end{figure}

As expected, the simulations demonstrate, in Fig. \ref{fig3}, that the gap
soliton is expelled by the trapping potential: starting from the initial
position, $X(t=0)$, the soliton moves with $X(t)=0.1\cosh (\lambda t)$ with $%
\lambda =0.00877$. Comparing this with the equation of motion for a particle
in potential $(1/2)kx^{2}$, we conclude that its effective mass is $m_{%
\mathrm{eff}}=-k/\lambda ^{2}\approx -1.30$, cf. the positive effective mass
(\ref{meff}) of the gap soliton in the case of the attractive nonlinearity.

\subsection{An exact solution for the linear system}

\begin{figure}[h]
\begin{center}
\includegraphics[height=4.cm]{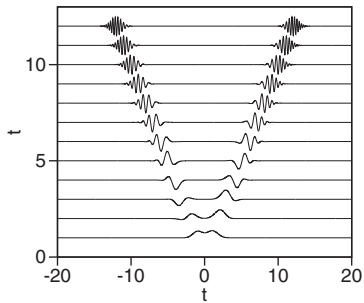}
\end{center}
\caption{Time evolution of $\mathrm{Re}\left( \Phi _{+}\right) $, as
obtained from numerical solution of Eq. (\protect\ref{1d2}) with $g=\protect%
\gamma =\Omega =0$, $k=0.125$, and initial conditions (\protect\ref{initial}%
). It is identical to the analytical solution given by Eqs. (\protect\ref%
{analyt})-(\protect\ref{initial}).}
\label{fig4}
\end{figure}

To complete the discussion of the 1D system placed in the trapping
potential, we note that it admits an exact solution in the linear limit, $%
g=\gamma =0$, if the Zeeman splitting is absent too, $\Omega =0$. Indeed, in
this case the solution of Eq. (\ref{1d2}) is easily found as%
\begin{eqnarray}
\Phi _{+}\left( x,t\right) &=&\frac{1}{2}\left[ \exp \left( \frac{i}{6}%
kx^{3}\right) \phi _{1}^{(\mathrm{arb})}\left( x+t\right) +\exp \left( -%
\frac{i}{6}kx^{3}\right) \phi _{2}^{(\mathrm{arb})}\left( x-t\right) \right]
,  \notag \\
\Phi _{-}\left( x,t\right) &=&\frac{i}{2}\left[ \exp \left( \frac{i}{6}%
kx^{3}\right) \phi _{1}^{(\mathrm{arb})}\left( x+t\right) -\exp \left( -%
\frac{i}{6}kx^{3}\right) \phi _{2}^{(\mathrm{arb})}\left( x-t\right) \right]
,  \label{analyt}
\end{eqnarray}%
where $\phi _{1,2}^{(\mathrm{arb})}\left( x\pm t\right) $ are arbitrary
complex functions of their arguments, which are determined if initial
conditions, $\Phi _{\pm }\left( x,t=0\right) $, are specified:%
\begin{gather}
\phi _{1}^{(\mathrm{arb})}(x)=\exp \left( -\frac{i}{6}kx^{3}\right) \left[
\Phi _{+}(x,t=0)-i\Phi _{-}(x,t=0)\right] ,  \notag \\
\phi _{2}^{(\mathrm{arb})}(x)=\exp \left( \frac{i}{6}kx^{3}\right) \left[
\Phi _{+}(x,t=0)+i\Phi _{-}(x,t=0)\right] .  \label{arb}
\end{gather}%
In particular, the numerical solution of Eq. (\ref{1d2}) with $g=\gamma
=\Omega =0$ and $k=0.125$, generated by initial conditions%
\begin{equation}
\Phi _{+}\left( x,t=0\right) =\Phi _{-}\left( x,t=0\right) =\exp \left(
-x^{2}\right) ,  \label{initial}
\end{equation}%
displayed in Fig. \ref{fig4}, exactly corresponds to the analytical solution
given by equations (\ref{analyt}) and (\ref{arb}). That is, the Gaussian
input splits into two wavelets traveling with speeds $\pm 1$, whose motion
is not affected by the trapping potential.

\section{Two-dimensional gap solitons in the free space}

\begin{figure}[h]
\begin{center}
\includegraphics[height=4.cm]{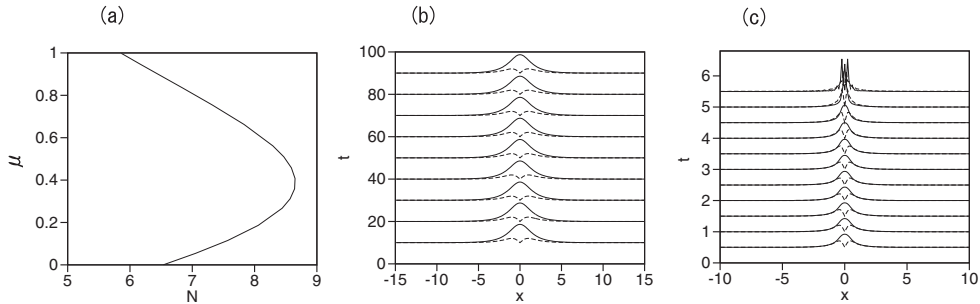}
\end{center}
\caption{(a) Chemical potential $\protect\mu $ vs. norm\ $N$ of the family
of the 2D gap solitons based on the semi-vortex ansatz (\protect\ref{semi}),
as obtained from the numerical solution of Eqs. (\protect\ref{U3}) and (%
\protect\ref{U4}) with $g=\Omega =1$ and $\protect\gamma =0$. Here and in
Fig. \protect\ref{fig6}(a) below, curves $\protect\mu (N)$ are not shown for
$-1<\protect\mu <0$, as in that interval they represent definitely unstable
solitons, which do not satisfy the Vakhitov-Kolokolov criterion, $d\protect%
\mu /dN<0$. (b) Stable evolution of components $|\Phi _{+}(x,y)|$ and $|\Phi
_{-}(x,y)|$ (solid and dashed lines, respectively) of the 2D soliton with $%
\protect\mu =0.796$, shown in cross section $y=0$. (c) The same as in (b),
but for an unstable 2D soliton with $\protect\mu =0.2672$.}
\label{fig5}
\end{figure}

In terms of polar coordinates $\left( r,\theta \right) $, soliton solutions
to the 2D system (\ref{2d}) can be looked for in the form of the semi-vortex
ansatz, which has the same general form as in Refs. \cite{Ben Li} and \cite%
{Sherman} (where, unlike the present analysis, the GPE system included the
kinetic-energy terms):
\begin{equation}
\Phi _{+}(r,\theta ,t)=e^{-i\mu t}\phi _{+}(r),~\Phi _{-}(r,\theta
,t)=e^{-\mu t}e^{i\theta }\phi _{-}(r).  \label{semi}
\end{equation}%
The substitution of this ansatz in Eq. (\ref{2d}) leads to the system of
radial equations for real amplitudes $\phi _{\pm }$:
\begin{eqnarray}
\frac{d\phi _{-}}{dr} &=&\mu \phi _{+}-\frac{\phi _{-}}{r}-\Omega \phi
_{+}+(g\phi _{+}^{2}+\gamma \phi _{-}^{2})\phi _{+},  \label{U3} \\
\frac{d\phi _{+}}{dr} &=&-\mu \phi _{-}-\Omega \phi _{-}-(g\phi
_{-}^{2}+\gamma \phi _{+}^{2})\phi _{-}.  \label{U4}
\end{eqnarray}%
Localized solutions to Eqs. (\ref{U3}), (\ref{U4}) can be readily obtained
by means of the shooting method. Here, we use scaling to fix $g=\Omega =1$.
The family of semi-vortex solitons is characterized by the dependence of $%
\mu $ on the 2D norm,
\begin{equation}
N=2\pi \int_{0}^{\infty }(|\phi _{+}|^{2}+|\phi _{-}|^{2})rdr,
\end{equation}%
which is displayed in Fig. \ref{fig5}(a) in the absence of the
cross-nonlinearity, $\gamma =0$.

The branch of the semi-vortex family with $d\mu /dN<0$ in Fig. \ref{fig5}(a)
is expected to be stable, while the other one, with $d\mu /dN>0$, should be
definitely unstable, according to the VK criterion. To check this
expectation, we have performed systematic simulations of the perturbed
evolution in the framework of Eq. (\ref{2d}). The simulations were run in
the 2D domain of extension $30\times 30$ in the present scaled units, using
the mesh size $256\times 256$ and the Runge-Kutta algorithm for marching
forward in time.

Results of the simulations completely corroborate the stability of the
branch which meets the VK criterion, even if, strictly speaking, it does not
produce a sufficient stability condition. As a typical example, Fig. \ref%
{fig5}(b) demonstrates the evolution of$\ |\Phi _{+}|$ and $|\Phi _{-}|$ for
$N=7.075$ and $\mu =0.796$, which is a semi-vortex soliton with amplitude $%
\phi _{+}(r=0)=1$, see Eq. (\ref{semi}). On the other hand, Fig. \ref{fig5}%
(c) confirms the instability of the soliton with $N=8.443$, $\mu =0.2672$,
and $\phi _{+}(r=0)=2.1$, which belongs to the lower branch in \ref{fig5}%
(a), with the positive slope, $d\mu /dN>0$. This unstable soliton (as well
as others belonging to the same branch) eventually suffers blowup
(collapse), which is a typical outcome of the evolution of unstable 2D
solitons in systems with the cubic self-attraction \cite{VK-review}.

\begin{figure}[h]
\begin{center}
\includegraphics[height=3.5cm]{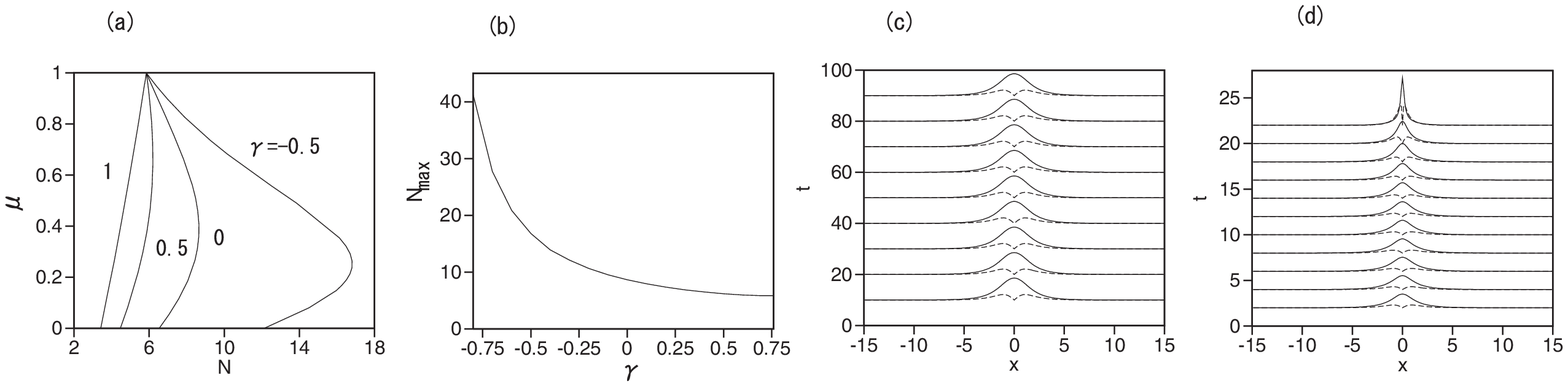}
\end{center}
\caption{(a) Curves $\protect\mu (N)$ for families of 2D\ gap solitons of
the semi-vortex type for $g=\Omega =1$ and different values of the relative
strength of the cubic interaction between the two components of the spinor
wave function: $\protect\gamma =1$, $0.5$, $0$, and $-0.5$. Note that
VK-stable segments of the $\protect\mu (N)$ curves, with the negative slope,
do not exist at $\protect\gamma >\protect\gamma _{\max }$, see Eq. (\protect
\ref{max}). (b) The largest values of the norm from the curves in panel (a),
$N_{\max }$, vs. $\protect\gamma $. (c) The evolution of $|\Phi _{+}(x,y)|$
and $|\Phi _{-}(x,y)|$ (solid and dashed lines, respectively), shown in
cross section $y=0$, for a stable gap soliton, with $\protect\mu =0.784$, at
$\protect\gamma =-0.5$. (d) The same, but for an unstable soliton, with $%
\protect\mu =0.8135$, at $\protect\gamma =1$.}
\label{fig6}
\end{figure}

As shown in Refs. \cite{Ben Li}, \cite{Sherman}, and \cite{new}, the
interplay of the nonlinear cross- and self-interactions in the 2D SO-coupled
system has a strong effect of the existence and stability of solitons. To
address this point, in Fig. \ref{fig6}(a) we display a set of $\mu (N)$
curves for families of gap semi-vortex solitons with fixed $g=\Omega =1$ (as
mentioned above) and different relative strengths of the cross-interaction, $%
\gamma =1$, $0.5$, $0$, and $-0.5$ (recall that $\gamma <0$ corresponding to
the repulsive sign of the cross interaction). For the sake of completeness,
the set includes, as a particular case, the curve for $\gamma =0$ from Fig. %
\ref{fig5}(a). An important fact is that VK-stable segments, with $d\mu /dN<0
$, do not exist at
\begin{equation}
\gamma >\gamma _{\max }\approx 0.77,  \label{max}
\end{equation}%
hence all the 2D gap solitons are \emph{completely unstable} in this case.
This conclusion agrees with the result reported in Ref. \cite{rf3}, where it
was found that all the 2D gap solitons in the system with the local
interactions are unstable for $\gamma =1$.

Thus, each $\mu (N)$ curve, plotted for $\gamma <\gamma _{\max }$, features
point $N_{\max }$ which is the boundary between the VK-stable and unstable
branches, with $d\mu /dN<0$ and $d\mu /dN>0$, respectively. The dependence
of $N_{\max }$ on $\gamma $ is displayed in Fig. \ref{fig6}(b). It is worthy
to note that $N_{\max }$ increases indefinitely at $\gamma \rightarrow -1$.
This property is explained by the fact that, at $\gamma =-1$, the
self-attractive and cross-repulsive interactions with equal strengths ($%
g=-\gamma =1$) may nearly cancel each other, thus helping to create gap
solitons at arbitrarily large $N$.

Another feature revealed by Fig. \ref{fig6}(a) is that all the $\mu (N)$
curves originate, at $\mu =1$, from a single point, with the norm
corresponding to the Townes' soliton, which is the fundamental solution of
the single 2D GPE with the cubic self-attraction and normal kinetic-energy
term \cite{VK-review}: $N=N_{\mathrm{Townes}}\approx 5.85$. The same is the
smallest value of $N_{\max }$ attained at $\gamma \rightarrow \gamma _{\max
} $, as can be seen in Fig. \ref{fig6}(b). This feature can be explained by
noting that, close to the top edge of the bandgap, i.e., at $\mu \rightarrow
1$ (recall we are using normalization $\Omega =1$), Eq. (\ref{U4}) yields a
small component $\phi _{-}$, which may be approximately eliminated in favor
of the larger one (cf. a similar approach developed, in a different context,
in Ref. \cite{Sherman}): $\phi _{-}\approx -\left( 1/2\right) d\phi _{+}/dr$%
. Substituting this in Eq. (\ref{U4}), we arrive at the single GPE in which
the effective kinetic-energy term is effectively restored:%
\begin{equation}
(\mu -1)\phi _{+}=-\frac{1}{2}\left( \frac{d^{2}}{dr^{2}}+\frac{1}{r}\frac{d%
}{dr}\right) \phi _{+}-\phi _{+}^{3}.  \label{Townes}
\end{equation}%
It is well known that Eq. (\ref{Townes}) generates a family of 2D Townes'
solitons, all having the same value of the norm, $N=N_{\mathrm{Townes}}$,
which explains that all the branches originate from this value of $N$ in
Fig. \ref{fig6}(a).

Lastly, Figs. \ref{fig6}(c) and (d) illustrate, severally, the stability and
instability of the gap solitons at $\gamma \neq 0$ [cf. the examples shown
for $\gamma =0$ in Figs. \ref{fig5}(b) and (b), respectively]: the stable
soliton is produced with $N=7.075$, $\mu =0.784$, and amplitude $\phi
_{+}(r=0)=1$ for $\gamma =-0.5$, while the unstable one, which eventually
blows up, is obtained with $N=5.45$, $\mu =0.8135$, and the same amplitude, $%
\phi _{+}(r=0)=1$, at $\gamma =1$.

\begin{figure}[h]
\begin{center}
\includegraphics[height=4.cm]{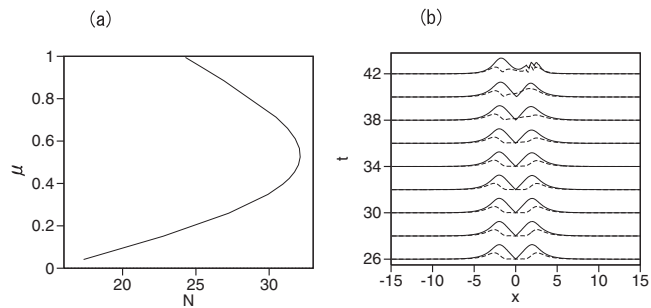}
\end{center}
\caption{(a) The $\protect\mu (N)$ curve for the family of excited states,
based on Eq. (\protect\ref{m}), with $g=\Omega =1$, $\protect\gamma =0$, and
$m=1$. (b) Unstable evolution of $|\Phi _{+}(x,y)|$ and $|\Phi _{-}(x,y)|$
(solid and dashed lines, respectively), shown in cross section $y=0$, of a
typical excited state with $m=1$.}
\label{fig7}
\end{figure}

The semi-vortices based on ansatz (\ref{semi}) represent fundamental gap
solitons. Following Ref. \cite{Ben Li}, it is possible to introduce a more
general ansatz, also compatible with system (\ref{2d}) of the SO-coupled
GPEs, for \textit{excited states} of the solitons, which are obtained by
adding common integer vorticity $m$ to both components of the spinor wave
function:
\begin{equation}
\Phi _{+}=e^{-i\mu t}e^{im\theta }\phi _{+}^{(m)},~\Phi _{-}=e^{-\mu
t}e^{i(m+1)\theta }\phi _{-}^{(m)}(r),  \label{m}
\end{equation}%
with real functions $\phi _{\pm }^{(m)}$ satisfying a system of radial
equations:
\begin{eqnarray}
\frac{d\phi _{-}^{(m)}}{dr} &=&\mu \phi _{+}^{(m)}-\frac{m\phi _{-}^{(m)}}{r}%
-\Omega \phi _{+}^{(m)}+\left[ g\left( \phi _{+}^{(m)}\right) ^{2}+\gamma
\left( \phi _{-}^{(m)}\right) ^{2}\right] \phi _{+}^{(m)},  \label{U5} \\
\frac{d\phi _{+}^{(m)}}{dr} &=&-\mu \phi _{-}^{(m)}+\frac{(m+1)\phi
_{+}^{(m)}}{r}-\Omega \phi _{-}^{(m)}-\left[ g\left( \phi _{-}^{(m)}\right)
^{2}+\gamma \left( \phi _{+}^{(m)}\right) ^{2}\right] \phi _{-}^{(m)},
\label{U6}
\end{eqnarray}%
cf. Eqs. (\ref{semi}) and (\ref{U3}), (\ref{U4}). Localized solutions to
Eqs. (\ref{U5}) and (\ref{U6}) can be again obtained with the help of the
shooting method. Figure \ref{fig7}(a) displays the $\mu (N)$ curve for the
family of the excited states with $m=1$ and $\gamma =0$.

However, unlike the fundamental semi-vortex solitons, all the excited states
are found to be unstable in direct simulations, although their $\mu (N)$
curve still demonstrates, in Fig. \ref{fig7}(a), a segment satisfying the VK
criterion (this fact stresses that the criterion is a necessary but not
sufficient stability condition). A typical example of the instability onset
is displayed in Fig. \ref{fig7}(b) for an excited state with $m=1$, $N=31.25$%
, $\mu =0.66$, whose stationary form is characterized by slope $d\phi
_{+}/dr=0.6$ at $r=0$. The instability spontaneously breaks the soliton's
symmetry and eventually leads to the collapse. Note that all the excited
states of semi-vortex solitons were also found to be unstable in the system
considered in Ref. \cite{Ben Li}, which included the kinetic-energy terms
and did not include the Zeeman splitting.

\section{Conclusion}

In this work we have demonstrated a novel mechanism for the formation of
stable 2D and 1D solitons in the free space, built by the SO (spin-orbit)
coupling combined with the sufficiently strong Zeeman splitting. While we
consider the SO coupling of the Rashba type, it is natural to expect that
more general forms of the coupling, such a Rashba-Dresselhaus \cite%
{Dresselhaus} mixture, will produce similar results, cf. Ref. \cite{Sherman}%
, which analyzed such a situation for ordinary solitons in the 2D system
including the kinetic-energy terms. Following recent work \cite{rf3}, we
consider the limit case when the energy of the SO coupling in the 2D system
is much larger than the kinetic energy, making it possible to neglect terms
accounting for the latter in the system of coupled GPEs (Gross-Pitaevskii
equations). In the combination with the Zeeman splitting, the system changes
its linear spectrum, opening a finite bandgap in it. In Ref. \cite{rf3}, it
was recently demonstrated that, in the case of equal strengths of the self-
and cross interactions between two components of the spinor wave function,
the cubic contact nonlinearity fails to produce stable 2D gap solitons in
the bandgap, therefore the analysis presented in Ref. \cite{rf3} relied upon
the use of long-range dipole-dipole interactions. The main finding reported
here is that the contact interactions of either sign, attractive or
repulsive, create \emph{stable} 2D fundamental gap solitons of the
semi-vortex type (with vorticities $0$ and $1$ in the two components),
provided that the strength of the cross interaction, normalized to that of
the self-interaction, does not exceed the critical value, $\approx 0.77$.
The boundary between stable and unstable subfamilies of 2D gap solitons in
the bandgap is exactly predicted by the VK (Vakhitov-Kolokolov) criterion.
Excited states of the semi-vortex solitons, created by adding equal
vorticities to each component, are found too in the bandgap, but they are
completely unstable. The analysis has also been performed for the 1D
reduction of the system. In that case, the family of fundamental gap
solitons, completely filling the bandgap, was found in the exact analytical
form, the entire family being stable. In the 1D system, motion of the gap
soliton in the trapping potential was studied too. The sign of the soliton's
effective mass is positive or negative, respectively, in the case of the
attractive or repulsive sign of the cubic nonlinearity. In particular, the
trapping potential was shown to be expulsive for the gap solitons with the
negative dynamical mass.

This work may be extended by considering moving 2D and 1D solitons and
collisions between them, as suggested by Eqs. (\ref{moving}) and (\ref%
{reduced}). In particular, it is interesting to find out how the motion
affects the stability of the 2D gap solitons.

\section*{Acknowledgments}

The work of B.A.M. is supported, in part, by the joint program in physics
between NSF and Binational (US-Israel) Science Foundation through project
No. 2015616, and by the Israel Science Foundation, through grant No. 1287/17.


\begin{thebibliography}{99}
\bibitem{SOC1} Y.-J. Lin, K. Jim\'{e}nez-Garc\'{\i}a, I. B. Spielman,
Spin-orbit-coupled Bose-Einstein condensates, Nature (London), \textbf{471},
83 (2011).

\bibitem{SOC2} D. L. Campbell, G. Juzeli\={u}nas, and I. B. Spielman,
Realistic Rashba and Dresselhaus spin-orbit coupling for neutral atoms,
Phys. Rev. A \textbf{84}, 025602 (2011); Y. Zhang, L. Mao, and C. Zhang,
Mean-Field Dynamics of spin-orbit coupled Bose-Einstein condensates, Phys.
Rev. Lett. \textbf{108}, 035302 (2012).

\bibitem{SOC-2D} Z. Wu, L. Zhang, W. Sun, X.-T. Xu, B.-Z. Wang, S.-C. Ji, Y.
Deng, S. Chen, X.-J. Liu, and J.-W. Pan, Realization of two-dimensional
spin-orbit coupling for Bose-Einstein condensates, Science \textbf{354},
83-88 (2016).

\bibitem{SOC-review} V. Galitski and I. B. Spielman, Spin-orbit coupling in
quantum gases, Nature \textbf{494}, 49-54 (2013); H. Zhai, Degenerate
quantum gases with spin-orbit coupling: a review, Rep. Progr. Phys. \textbf{%
78}, 026001 (2015).

\bibitem{SOC-vortices} T. Kawakami, T. Mizushima, and K. Machida, Textures
of \textit{F =} 2 spinor Bose-Einstein condensates with spin-orbit coupling,
Phys. Rev. A 84, 011607 (2011); H. Sakaguchi and B. Li, Vortex lattice
solutions to the Gross-Pitaevskii equation with spin-orbit coupling in
optical lattices, Phys. Rev. A 87, 015602 (2013).

\bibitem{SOC-monopole} G. J. Conduit, Line of Dirac monopoles embedded in a
Bose-Einstein condensate, Phys. Rev. A \textbf{86}, 021605(R) (2012).

\bibitem{SOC-skyrmion} C. J. Wu, I. Mondragon-Shem, and X.-F. Zhou,
Unconventional Bose-Einstein condensations from spin-orbit coupling, Chin.
Phys. Lett. \textbf{28}, 097102 (2011); T. Kawakami, T. Mizushima, M. Nitta,
and K. Machida, Stable skyrmions in \textit{SU}(2) gauged Bose-Einstein
condensates, Phys. Rev. Lett. 109, 015301 (2012).

\bibitem{soliton-1D} Y. Xu, Y. Zhang, and B. Wu, Bright solitons in
spin-orbit-coupled Bose-Einstein condensates, Phys. Rev. A \textbf{87},
013614 (2013); V. Achilleos, D. J. Frantzeskakis, P. G. Kevrekidis, and D.
E. Pelinovsky, Matter-wave bright solitons in spin-orbit coupled
Bose-Einstein condensates, Phys. Rev. Lett. \textbf{110}, 264101 (2013); Y.
V. Kartashov, V. V. Konotop, and F. Kh. Abdullaev, Gap Solitons in a
Spin-Orbit-Coupled Bose-Einstein Condensate, \textit{ibid}. \textbf{111},
060402 (2013); L. Salasnich and B. A. Malomed, Localized modes in dense
repulsive and attractive Bose-Einstein condensates with spin-orbit and Rabi
couplings, Phys. Rev. A \textbf{87}, 063625 (2013); L. Wen, Q. Sun, Y. Chen,
D.-S. Wang, J. Hu, H. Chen, W.-M. Liu, G. Juzeli\={u}nas, B. A. Malomed, and
A.-C. Ji, Motion of solitons in one-dimensional spin-orbit-coupled
Bose-Einstein condensates, Phys. Rev. A \textbf{94}, 061602(R) (2016); Y. V.
Kartashov and V. V. Konotop, Solitons in Bose-Einstein condensates with
helicoidal spin-orbit coupling, Phys. Rev. Lett. \textbf{118}, 190401 (2017).

\bibitem{Brand} O. Fialko, J. Brand, and U. Z\"{u}licke, Soliton
magnetization dynamics in spin-orbit-coupled Bose-Einstein condensates,
Phys. Rev. A \textbf{85}, 051605(R) (2012).

\bibitem{Ben Li} H. Sakaguchi, B. Li, and B. A. Malomed, Creation of
two-dimensional composite solitons in spin-orbit-coupled self-attractive
Bose-Einstein condensates in free space, Phys. Rev. E \textbf{89}, 032920
(2014).

\bibitem{soliton-2D} V. E. Lobanov, Y. V. Kartashov, and V. V. Konotop,
Fundamental, multipole, and half-vortex gap solitons in spin-orbit coupled
Bose-Einstein condensates, Phys. Rev. Lett. \textbf{112}, 180403 (2014); L.
Salasnich, W. B. Cardoso, and B. A. Malomed, Localized modes in
quasi-two-dimensional Bose-Einstein condensates with spin-orbit and Rabi
couplings, Phys. Rev. A \textbf{90}, 033629 (2014).

\bibitem{Han Pu} Y.-C. Zhang, Z.-W. Zhou, B. A. Malomed, and H. Pu, Stable
solitons in three-dimensional free space without the ground state:
Self-trapped Bose-Einstein condensates with spin-orbit coupling, Phys. Rev.
Lett. \textbf{115}, 253902 (2015).

\bibitem{Sherman} H.~Sakaguchi, E.~Y.~Sherman, and B.~A.~Malomed, Vortex
solitons in two-dimensional spin-orbit coupled Bose-Einstein condensates:
Effects of the Rashba-Dresselhaus coupling and the Zeeman splitting, Phys.
Rev. E \textbf{94}, 032202 (2016).

\bibitem{old-review} B. A. Malomed, D. Mihalache, F. Wise, and L. Torner,
Spatiotemporal optical solitons. J. Optics B: Quant. Semicl. Opt. \textbf{7}%
, R53-R72 (2005); Viewpoint: On multidimensional solitons and their legacy
in contemporary Atomic, Molecular and Optical physics, J. Phys. B: At. Mol.
Opt. Phys. \textbf{49}, 170502 (2016).

\bibitem{Dumitru} D. Mihalache, Linear and nonlinear light bullets: Recent
theoretical and experimental studies, Rom. J. Phys. \textbf{57}, 352 (2012)

\bibitem{recent-review} B. A. Malomed, Multidimensional solitons:
Well-established results and novel findings, Eur. Phys. J. Special Topics
\textbf{225}, 2507-2532 (2016).

\bibitem{Santos} S. Sinha, R. Nath, and L. Santos, Trapped two-dimensional
condensates with synthetic spin-orbit coupling, Phys. Rev. Lett. \textbf{107}%
, 270401 (2011).

\bibitem{Drummond-Pu} B. Ramachandhran, B. Opanchuk, X.-J. Liu, H. Pu, P. D.
Drummond, and H. Hu, Half-quantum vortex state in a spin-orbit-coupled
Bose-Einstein condensate, Phys. Rev. A \textbf{85}, 023606 (2012).

\bibitem{rf3} Y.~Li, Y.~Liu, Z.~Fan. W.~Pang, S.~Fu, and B. A. Malomed,
Two-dimensional dipolar gap solitons in free space with spin-orbit coupling,
Phys. Rev. A \textbf{95}, 063613 (2017).

\bibitem{Rashba} Y. A. Bychkov and E. I. Rashba, Oscillatory effects and the
magnetic susceptibility of carriers in inversion layers, J. Phys. C \textbf{%
17}, 6039 (1984); E. I. Rashba and E. Y. Sherman, Spin orbital band
splitting in symmetric quantum wells, Phys. Lett. A \textbf{129}, 175-179
(1988).

\bibitem{Zeeman} D. L. Campbell, G. Juzeli\={u}nas, and I. B. Spielman,
Realistic Rashba and Dresselhaus spin-orbit coupling for neutral atoms,
Phys. Rev. A \textbf{84}, 025602 (2011).

\bibitem{Sterke} A. B. Aceves and S. Wabnitz, Self-induced transparency
solitons in nonlinear refractive periodic media, Phys. Lett. A\textbf{\ 141}%
, 37-42 (1989); D. N. Christodoulides and R. I. Joseph, Slow Bragg solitons
in nonlinear periodic structures, Phys. Rev. Lett. \textbf{62}, 1746-1748
(1989); C. M. de Sterke and J. E. Sipe, Gap solitons, Progr. Optics \textbf{%
33}, 203-260 (1994).

\bibitem{Krug} B. J. Eggleton, R. E. Slusher, C. M. de Sterke, P. A. Krug,
and J. E. Sipe, Bragg Grating solitons, Phys. Rev. Lett. \textbf{76},
1627-1630 (1996).

\bibitem{plasmon} E. A. Ostrovskaya, J. Abdullaev, M. D. Fraser, A. S.
Desyatnikov, and Yu. S. Kivshar, Self-localization of polariton condensates
in periodic potentials, Phys. Rev. Lett. \textbf{110}, 170407 (2013); E. A.
Cerda-M\'{e}ndez, D. Sarkar, D. N. Krizhanovskii, S. S. Gavrilov, K.
Biermann, M. S. Skolnick, and P. V. Santos, Exciton-polariton gap solitons
in two-dimensional lattices, \textit{ibid}. \textbf{111}, 146401 (2013).

\bibitem{BEC} V. A. Brazhnyi and V. V. Konotop, Theory of nonlinear matter
waves in optical lattices, Mod. Phys. Lett. B \textbf{18}, 627 (1994); O.
Morsch and M. Oberthaler, Dynamics of Bose-Einstein condensates in optical
lattices, Rev. Mod. Phys. \textbf{78}, 179 (2006).

\bibitem{Oberthaler} B. Eiermann, T. Anker, M. Albiez, M. Taglieber, P.
Treutlein, K. P. Marzlin, and M. K. Oberthaler, Bright Bose-Einstein gap
solitons of atoms with repulsive interaction, Phys. Rev. Lett. \textbf{92},
230401 (2004).

\bibitem{flatband} H.-Y. Hui, Y. Zhang, C. Zhang, and V. W. Scarola,
Superfluidity in the absence of kinetics in spin-orbit-coupled optical
lattices, Phys. Rev. A \textbf{95}, 033603 (2017).

\bibitem{Ho} T.-L. Ho, Spinor Bose condensates in optical traps, Phys. Rev.
Lett. \textbf{81}, 742-745 (1998).

\bibitem{Cr} Y. Deng, J. Cheng, H. Jing, C.-P. Sun, and S. Yi,
Spin-orbit-coupled dipolar Bose-Einstein condensates, Phys. Rev. Lett.
\textbf{108}, 125301 (2012).

\bibitem{Xunda} R. M. Wilson, B. M. Anderson, and C. W. Clark, Meron ground
state of Rashba Spin-Orbit-Coupled dipolar bosons, Phys. Rev. Lett. \textbf{%
111}, 185303 (2013); S. C. Ji, L. Zhang, X. T. Xu, Z. Wu, Y. J. Deng, S.
Chen, and J. W. Pan, Softening of roton and phonon modes in a Bose-Einstein
condensate with spin-orbit coupling, \textit{ibid}. \textbf{114}, 105301
(2015); Y. Xu, Y. Zhang, and C. Zhang, Bright solitons in a 2D
spin-orbit-coupled dipolar Bose-Einstein condensate, Phys. Rev. A \textbf{92}%
, 013633 (2015); X. Jiang, Z. Fan, Z. Chen, W. Pang, Y. Li, and B. A.
Malomed, Two-dimensional solitons in dipolar Bose-Einstein condensates with
spin-orbit coupling, \textit{ibid}. \textbf{93}, 023633 (2016); T. Oshima
and Y. Kawaguchi, Spin Hall effect in a spinor dipolar Bose-Einstein
condensate, \textit{ibid}. \textbf{93}, 053605 (2016).

\bibitem{Tikhonenkov} I. Tikhonenkov, B. A. Malomed, and A. Vardi,
Anisotropic Solitons in Dipolar Bose-Einstein Condensates, Phys. Rev. Lett.
\textbf{100}, 090406 (2008); P. K\"{o}berle, D. Zajec, G. Wunner, and B. A.
Malomed, Creating two-dimensional bright solitons in dipolar Bose-Einstein
condensates, Phys. Rev. A \textbf{85}, 023630 (2012).

\bibitem{embedded} J. Yang, B. A. Malomed, and D. J. Kaup, Embedded solitons
in second-harmonic-generating systems, Phys. Rev. Lett. \textbf{83},
1958-1961 (1999); A. R. Champneys, B. A. Malomed, J. Yang, and D. J. Kaup,
``Embedded solitons" : solitary waves in resonance with the linear spectrum,
Physica D 152-153, 340-354 (2001); J. Yang, Fully localized two-dimensional
embedded solitons, Phys. Rev. A \textbf{82}, 053828 (2010).

\bibitem{Feshbach} S. B. Papp, J. M. Pino, and C. E. Wieman, Tunable
miscibility in a dual-species Bose-Einstein condensate, Phys. Rev. Lett.
\textbf{101}, 040402 (2008); P. Zhang, P. Naidon, and M. Ueda, Independent
Control of Scattering Lengths in Multicomponent Quantum Gases, \textit{ibid}%
. \textbf{103}, 133202 (2009); F. Wang, X. Li, D. Xiong, and D. Wang, A
double species $^{23}$Na and $^{87}$Rb Bose-Einstein condensate with tunable
miscibility via an interspecies Feshbach resonance, J. Phys. B: At. Mol.
Opt. Phys. \textbf{49}, 015302 (2016).

\bibitem{Giovanazzi} S. Giovanazzi, A. G\"{o}rlitz, and T. Pfau, Tuning the
dipolar interaction in quantum gases, Phys. Rev. Lett. \textbf{89}, 130401
(2002).

\bibitem{VK} M. Vakhitov and A. Kolokolov, Stationary solutions of the wave
equation in a medium with nonlinearity saturation, Radiophys. Quantum
Electron. \textbf{16}, 783 (1973).

\bibitem{VK-review} L. Berg\'{e}, Wave collapse in physics: principles and
applications to light and plasma waves, Phys. Rep. \textbf{303}, 259 (1998);
G. Fibich, \textit{The Nonlinear Schr\"{o}dinger Equation: Singular
Solutions and Optical Collapse} (Springer: Cham, 2015).

\bibitem{Goldstone} V. Gurarie and J. T. Chalker, Bosonic excitations in
random media, Phys. Rev. B \textbf{68}, 134207 (2003).

\bibitem{Weyl} O. Vafek and A. Vishwanath, Dirac fermions in solids: From
high-\textit{T}$_{\mathrm{c}}$ cuprates and graphene to topological
insulators and Weyl semimetals, Ann. Rev. Cond. Matt. \textbf{5}, 83-112
(2014).

\bibitem{Haddad1} L. H. Haddad and L. D. Carr, The nonlinear Dirac equation
in Bose Einstein condensates: Foundation and symmetries, Physica D, \textbf{%
238}, 1413 (2009); L. H. Haddad,C. M. Weaver, and L. D. Carr, The nonlinear
Dirac equation in Bose-Einstein condensates: I. Relativistic solitons in
armchair nanoribbon optical lattices, New J. Phys. 17, 063033 (2015); L. H.
Haddad and L. D. Carr, The nonlinear Dirac equation in Bose-Einstein
condensates: II. Relativistic soliton stability analysis, New J. Phys.
\textbf{17} 063034 (2015); M. J. Ablowitz, and Y. Zhu, Nonlinear waves in
shallow honeycomb lattices, SIAM J. Appl. Math. \textbf{72}, 240 (2012).

\bibitem{Haddad2} D. E. Pelinovsky and A. Stefanov, Asymptotic stability of
small gap solitons in nonlinear Dirac equations, J. Math. Phys. \textbf{53},
073705 (2012); L. H. Haddad and L. D. Carr, The nonlinear Dirac equation in
Bose-Einstein condensates: vortex solutions and spectra in a weak harmonic
trap, New J. Phys. \textbf{17}, 113011 (2015).

\bibitem{Kevrekidis} J. Cuevas-Maraver, P. G. Kevrekidis, A. Saxena, A.
Comech, and R. Lan, Stability of solitary waves and vortices in a 2D
nonlinear Dirac model, Phys. Rev. Lett. \textbf{116}, 214101 (2016).

\bibitem{Luca} L. Salasnich, A. Parola, and L. Reatto, Effective wave
equations for the dynamics of cigar-shaped and disk-shaped Bose condensates,
Phys. Rev. A \textbf{65}, 043614 (2002).

\bibitem{Maciek} A. Muryshev, G. V. Shlyapnikov, W. Ertmer, K. Sengstock,
and M. Lewenstein, Dynamics of dark solitons in elongated Bose-Einstein
condensates, Phys. Rev. Lett. \textbf{89}, 110401 (2002).

\bibitem{Luca2} L. Salasnich1 and B. A. Malomed, Vector solitons in nearly
one-dimensional Bose-Einstein condensates, Phys. Rev. A \textbf{74}, 053610
(2006).

\bibitem{we} H.~Sakaguchi and B.~A.~Malomed, One- and two-dimensional
solitons in $\mathcal{PT}$-symmetric systems emulating spin-orbit coupling,
New J. Phys. \textbf{18}, 105005 (2016).

\bibitem{we-old} H. Sakaguchi and B. A. Malomed. Dynamics of positive- and
negative-mass solitons in optical lattices and inverted traps, J. Phys. B
\textbf{37}, 1443-1459 (2004).

\bibitem{negative-mass} M. Salerno, V. V. Konotop, and Yu. V. Bludov,
Long-living Bloch oscillations of matter waves in periodic potentials, Phys.
Rev. Lett. \textbf{101}, 030405 (2008).

\bibitem{repulsion} O. Zobay, S. Potting, P. Meystre, and E. M. Wright,
Creation of gap solitons in Bose-Einstein condensates, Phys. Rev. A \textbf{%
59}, 643-648 (1999); P. J. Y. Louis, E. A. Ostrovskaya, C. M. Savage, and Y.
S. Kivshar, Bose-Einstein condensates in optical lattices: Band-gap
structure and solitons, \textit{ibid}. \textbf{67}, 013602 (2003); N. K.
Efremidis and D. N. Christodoulides, Lattice solitons in Bose-Einstein
condensates, \textit{ibid}. \textbf{67}, 063608 (2003).

\bibitem{new} Y. Li, Z. Luo, Y. Liu, Z. Chen, C. Huang, S. Fu, H. Tan, and
B. A. Malomed, Two-dimensional solitons and quantum droplets supported by
competing self- and cross-interactions in spin-orbit-coupled condensates,
New J. Phys., in press.

\bibitem{Dresselhaus} G. Dresselhaus, Spin-orbit coupling effects in zinc
blende structures, Phys. Rev. \textbf{100}, 580-586 (1955).
\end{thebibliography}
\end{document}